# Beta-plane turbulence: experiments with altimetry

Y. Zhang and Y. D. Afanasyev[a]

*Memorial University of Newfoundland, St. John's, Canada*

[a]Corresponding author address:

Yakov Afanasyev

Memorial University of Newfoundland, St. John's, NL, Canada

E-mail: afanai@mun.ca

http://www.physics.mun.ca/~yakov

**Abstract**

Results from a new series of experiments on barotropic and baroclinic turbulent flows generated in a rotating tank with topographic β-effect are presented. The velocity fields are measured by the Altimetric Imaging Velocimetry. The turbulent flows observed in the experiments develop zonal jets which are latent in a stationary forced-dissipative regime of the flow but become prominent in the decaying flow. The two-dimensional energy spectra of the flows exhibit the development of anisotropy towards zonal motions. The experiments demonstrate dual turbulent cascade with energy and enstrophy ranges. The frequency-wavenumber spectra reveal the presence of Rossby waves at low wavenumbers which are excited by the turbulent motions. The experimental results are compared with available theory of β-plane turbulence.

## I. INTRODUCTION

Two-dimensional β-plane turbulence is an important conceptual model of turbulent flows on rotating planets where the variation of the Coriolis parameter with latitude is of importance for motions of large scale. Before we consider the β-plane flows, we shall review briefly an even more basic concept of two-dimensional turbulence without β-effect (see review papers by Danilov and Gurarie[1] and Boffetta and Ecke[2]). A Kolmogorov type theory formulated by Kraichnan[3] predicts that a forced-dissipative two-dimensional turbulence develops a dual cascade, where energy is transferred from the scale where the forcing is applied to larger scales while enstrophy is transferred to smaller scales. The slopes of energy spectrum predicted by the theory are -5/3 in the energy range and -3 in the enstrophy range. Note that numerical simulations by Fontane et al.[4] performed at high Reynolds number showed that the theoretical -5/3 scaling law can be steepened to -2 when long-lived vortex cores develop in the flow. An extension beyond two-dimensionality is provided by rotating shallow water (RSW) theory (e.g. Zeitlin[5]) which takes into account vortex stretching, the effect of the finite radius of deformation and inertia-gravity waves. Recent laboratory experiments with RSW turbulence on the f-plane by Afanasyev and Craig[6] showed similarity to two-dimensional turbulence with the spectral slopes of approximately -5/3 and (somewhat steeper than) -3. These experimental results are in agreement with results of numerical simulations by Yuan and Hamilton[7]. An effect of asymmetry between anticyclonic and cyclonic vortices with the anticyclonic vortices prevailing was observed in the experiments in Ref. (6); this is one of the effects that make RSW turbulence different from its two-dimensional counterpart. In order to eliminate the β-effect, a cylindrical tank with a paraboloidal bottom of the same shape as that of the rotating water surface was employed in Ref. (6). Note that in the present work we use a similar apparatus but with an inverted paraboloidal bottom to increase β-effect rather than to eliminate it.

While two-dimensional and RSW turbulence are isotropic in the horizontal plane, the turbulence with the β-effect is not. Even when the forcing is isotropic, an anisotropy develops in the form of zonal jets. This phenomenon is of great interest for applications to atmospheric circulations on gas giants (see e.g. review by Dowling[8]) as well to recently discovered jets in the Earth's oceans (Maximenko et al.[9], Maximenko et al.[10] and Sokolov and Rintoul[11]). Note that while on Jupiter and Saturn jets are prominent features of the circulation, the oceanic jets are often "lost" in a much stronger eddy field and can be revealed only as a result of time averaging of the observed fields. In the spectral representation, the anisotropy of the flows manifests itself in the concentration of energy at the $k_x = 0$ axis in the wavenumber plane. Theory predicts that the zonal spectrum of the flow defined as $E_Z = E(k_x = 0, k_y)$ has a slope of -5 (Rhines[12], Sukoriansky et al.[13]). Here $E(k_x, k_y)$ is a two-dimensional energy spectrum which will be properly defined later in this article.

Baroclinic instability is a major underlying process which generates eddies in atmospheric and oceanic large-scale turbulent flows. At a mature stage of its development baroclinic instability becomes equilibrated such that the flow becomes barotropic to a significant degree. The barotropic component of the flow is coupled to the baroclinic one (e.g. Cushman-Roisin and Beckers[14]). Barotropization process allows one to use barotropic models for the investigation of turbulent flows. Observations of β-plane turbulence in numerical simulations of different authors suggested that barotropic flows tend to generate somewhat weaker jets than those in baroclinic flows. It seems that an energy source capable of sustaining persistently high energy level in the flow is an important factor of strong jet formation (e.g. Rhines[15] , Panetta[16]). In baroclinic flows (even when the dynamics is mainly barotropic) the energy source is provided by eddies slowly releasing their available potential energy.

Turbulent flows on the β-plane have been a subject of study in laboratory experiments by different authors. Zonal jets were observed in the experiment by Whitehead[17] where the flows were induced locally by a vertically oscillating disk. A formation of a zonal flow was also observed by Colin de Verdiere[18] who generated the flow by a periodic excitation of sources and sinks. Recent experiments by Afanasyev et al.[19] and Slavin and Afanasyev[20] showed that a pattern of zonal jets can be formed westward of the area where perturbations are located. The perturbations were created by buoyancy sources. It was shown that the jets formed via a β-plume mechanism. A β-plume is an almost zonal circulation consisting of two zonal jets flowing in opposite directions. The circulation is formed via radiation of Rossby waves by a localized perturbation (see e.g. Afanasyev et al.[19] , Stommel[21], and Davey and Killworth[22]). In the experiments in Ref. (19) eddies eventually filled a significant area of the domain. However, the jets were still prominent in this turbulent flow. While it is clear how a single localized perturbation can create jets to the west of its location, it is much less clear how eddies in a turbulent flow "cooperate" to sustain jets. Rather than using a forcing applied in a limited area of the domain, one can use a uniformly distributed forcing to create an eddy field in the entire domain. Afanasyev and Wells[23] used an electromagnetic (EM) forcing to generate a turbulent flow on the polar β-plane. In EM experiments the Lorentz force is created by using a combination of permanent magnets located at the bottom of the tank and the horizontal electric current flowing through the layer of conducting fluid. Experiments in Ref. (23) showed the formation of zonal flows in the forced-dissipative β-plane turbulence, but they were somewhat limited by the size of the apparatus. Further EM experiments in a setup similar to that used in Ref. (23) were performed recently by Espa et al.[24]. A different kind of distributed forcing was used by Read et al.[25] in their experiments on the Coriolis platform. The authors used convective forcing where the surface of the water in the tank was sprayed with saline water. More dense saline water formed small-scale convective plumes descending to the bottom thus creating small-scale turbulence. A system of zonal jets was observed in the experiments in Ref. (25). The slope of the turbulent energy spectrum was measured to be close to -5/3 in the energy range as predicted by theory. Since the forcing was at very small scale, the enstrophy range was not resolved in those experiments.

In what follows, we describe laboratory experiments on a rotating table with the (topographic) β-effect. Two experiments were performed; in the first experiment the forcing was barotropic and generated by the EM method while in the second experiment the forcing was baroclinic and generated by heating the fluid at the bottom of the tank. The purpose of these experiments was to investigate the properties of the flows including their spectral characteristics both in wavenumber space and in frequency-wavenumber space. We used a relatively large tank and an intermediate/small-scale forcing in order to observe the dual cascade including the energy and enstrophy intervals. Our optical altimetry system allowed us to measure fields with high spatial and temporal resolution which was necessary for spectral analyses. The dynamics of Rossby waves in the forced-dissipative β-plane turbulence is elucidated. Our experimental results show how idealized theory or numerical simulations hold in application to real flows.

In Sec. II of this paper we describe the setup of our apparatus as well as the optical altimetry technique used to measure the gradient of the surface elevation field, from which we obtain the velocity and vorticity fields. In Sec. III the results of the experiments and their analyses are reported. Concluding remarks are given in Sec. IV.

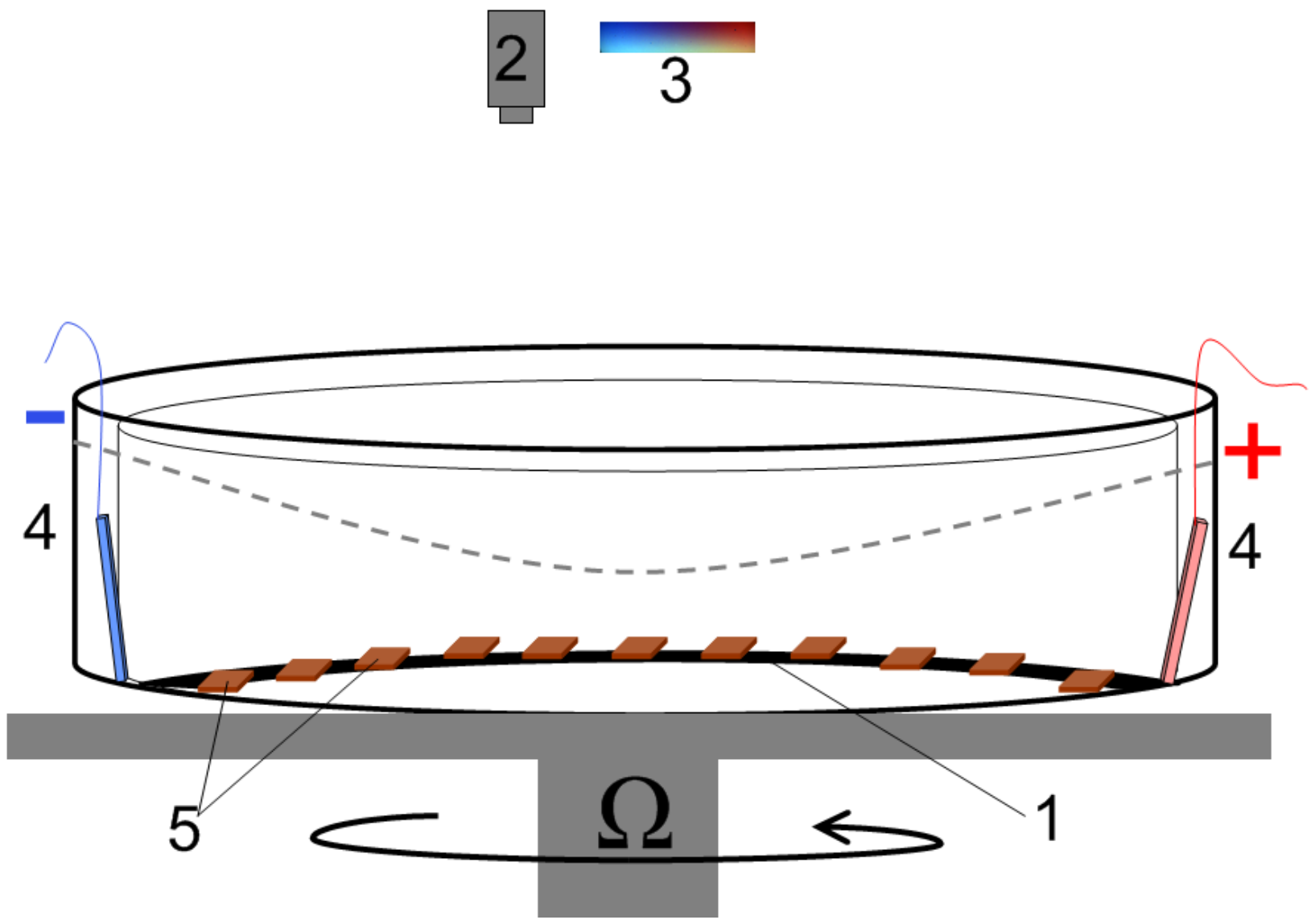

Fig.1. (Color online) Sketch of the experimental setup: rotating tank and the inner container with the paraboloid false bottom (1) filled with water, video camera (2), high brightness TFT panel displaying the color mask (3), electrodes (4) and permanent magnets (5). The dashed gray line shows the free water surface in a state of solid-body rotation.

## II. LABORATORY APPARATUS AND TECHNIQUE

The experiments were performed in a cylindrical tank of radius $R_t$ = 55 cm installed on a rotating table (Fig. 1). The tank was filled with water and rotated in an anticlockwise direction at a rate of $\Omega$ = 2.32 rad $s^{-1}$.

In the first experiment with an electromagnetic (EM) forcing an inner container in the form of a cylinder with a paraboloidal bottom was inserted into the tank and was concentric with the tank. The radius of the inner cylinder was $R$ = 45 cm. The bottom was an inverted paraboloid such that the bottom topography was $h_b = H_b - c_b r^2$ , where $H_b$ = 8 cm, $c_b = 4 \times 10^{-3}$ $cm^{-1}$ and $r$ is the radial distance. The flow was generated by an electromagnetic (EM) method. For this purpose the bottom of the inner container was fitted uniformly at $l_f$ = 4.6 cm intervals with about 300 neodymium magnets. The magnets were in a form of square tiles $l_m$ =2.5 cm wide. Vertical component of magnetic field of each magnet was of order of 1 Tesla. The poles of the magnets were oriented such that their polarity alternated between neighbouring magnets. Two electrodes made from chemically neutral material (graphite) were placed on the outside of the wall of the inner container to prevent bubbles forming on the electrodes to enter the area of interest. Small holes were drilled in the wall to allow electric current to flow through the fluid inside. A voltage of approximately 117 V DC was applied to the electrodes. The electrical conductivity of water was increased by adding NaCl in the amount of approximately 35 parts per thousand. The resulting electric current between the electrodes was about 15 A. A combination of the horizontal electric current with the vertical magnetic field results in the Lorentz force acting in the horizontal direction perpendicular to that of the electric current. The Lorentz force is a body force which acts on the fluid above each magnet. The effective vertical extent of the volume where the force is applied is approximately equal to the width of a magnet. The electromagnetic method provides an effective means of forcing the fluid in a controlled manner (for more details see Refs 26, 27).

The second experiment was performed in the tank without the cylindrical insert or the paraboloidal bottom. The flow was forced thermally using a heating wire at the bottom of the tank. The wire was fitted in an approximately uniform pattern such that the distance between the neighbouring segments of the wire was about 4.5 cm. The power provided by the heater was 2300 W.

The height $h$ of the water surface under a solid-body rotation varies quadratically with the distance $r$ from the axis of rotation

$$h(r) = H_0 + \frac{\Omega^2}{2g}\left(r^2 - \frac{R_t^2}{2}\right), \tag{1}$$

where $H_0$ is the depth of the layer in the absence of rotation and $g$ is the gravitational acceleration. The depth of the water in the tank is given by the difference between the height of the surface and bottom topography, $H = h - h_b$. In the EM experiment the water was quite shallow ($H$ = 2.5 cm) at the center of the tank and deep ($H$ = 14 cm) near the wall at radius $R$. In the heating wire experiment the bottom of the tank was flat ($h_b = 0$) and the water depth varied between 4 cm at the center and 12 cm near the wall ($r = R_t$) with the average depth $H_0$ = 8 cm.

The dynamical equivalence of the varying depth of the layer to the varying Coriolis parameter, results from the conservation of the potential vorticity (PV) defined as $q = (\zeta + f_0)/H$. Here $f_0 = 2\Omega$ is the Coriolis parameter and $\zeta$ is the vertical component of the relative vorticity. A local Cartesian coordinate system can be introduced at some reference radius $r_0$ such that a positive x-direction is to the East and a positive y-direction is to the North. Note that the North pole is at the center of the tank rotating anticlockwise. The β-plane with the linearly varying Coriolis parameter, $f = f_0 + \beta y$, can then be introduced by defining the β-parameter as

$$\beta = \frac{4\left(\Omega^2 / 2g + c_b\right)\Omega r_0}{H(r_0)}. \tag{2}$$

The Altimetric Imaging Velocimetry (AIV) system was used to observe perturbations of the surface and to measure two components of the gradient $\nabla\eta = (\partial\eta/\partial x,\ \partial\eta/\partial y)$ of the surface elevation $\eta$ in the horizontal plane ($x$, $y$). The AIV technique was described in detail in Ref. 28. The $\nabla\eta$ field was measured with a spatial resolution of approximately 2 vectors per millimeter such that a total size of the array was $1900 \times 1900$ and with a temporal resolution of 5 fields per second. The surface velocity of the flow can be determined from the measured gradient of surface elevation using shallow water and quasigeostrophic approximations which yields

$$\mathbf{U} = \frac{g}{f_0}\left(\mathbf{n}\times\nabla\eta\right) - \frac{g}{f_0^2}\frac{\partial}{\partial t}\nabla\eta - \frac{g^2}{f_0^3}J\left(\eta,\nabla\eta\right), \qquad (3)$$

where $\mathbf{U}$ is the horizontal velocity vector, $\mathbf{n}$ is the vertical unit vector and $J(A,B) = \frac{\partial A}{\partial x}\frac{\partial B}{\partial y} - \frac{\partial B}{\partial x}\frac{\partial A}{\partial y}$ is the Jacobian operator. The first term on the RHS of Eq. 3 is the geostrophic velocity. The second and third terms are due to transient and nonlinear effects and their relative importance is determined by the temporal Rossby number $Ro_T = 1/(f_0 T)$ and the Rossby number $Ro = U/(f_0 L)$ respectively. Here $T$ is the time scale of the flow evolution, while $U$ and $L$ are velocity and length scales of the flow.

## III. EXPERIMENTAL RESULTS AND ANALYSES

We performed two experiments with different forcing which will be referred as barotropic (Bt) experiment with EM forcing and baroclinic (Bc) experiment with thermal forcing. Note that in fact an ensemble of EM experiments with similar values of the dimensional control parameters was performed. The mean water depth and the forcing strength varied within approximately 15% between the experiments. Here we give the results which are typical for the ensemble. In what follows we discuss the Bt and Bc experiments in parallel emphasizing the similarities and differences between them.

### A. Observations of the flow evolution

Fig. 2 (enhanced online) shows a typical evolution of the flow in the forced regime and after the forcing stops in the Bt experiment. The initial period, shortly after the forcing is switched on, is shown in top row of Fig. 2. A regular array of vortices of alternating sign is formed such that there are approximately 10 vortices of the same sign across the tank. Note that initially the vortices between the adjacent magnets have double cores (Fig. 2 c) because the magnets generate vortex dipoles directed along parallel lines in the directions opposite to each other. However these cores rapidly coalesce into a single vortex. The flow is much stronger in the center of the tank where the water is shallow. The second row of Fig. 2 shows the flow at the end of the relatively long forcing period when the flow is at approximately steady state, the turbulence is fully developed; strong vortices are abundant. They are no longer attached to specific locations although some indications of the regular forcing array can be seen in the azimuthal velocity map (Fig. 2 e). The typical size of the vortices corresponds well to the forcing scale. The third row of Fig. 2 shows the flow shortly after the forcing is switched off. The vortices are still prominent, yet a certain alignment of vortices in the zonal direction is also evident. The map of the x-component of velocity, $u$, shown in Fig. 2 h indicates the emergence of zonal jets in the flow. Finally the bottom row of Fig. 2 shows the decaying flow at the very end of the experiment when vortices no longer dominate but zonal currents still present. Note an interesting spiral pattern of jets in this flow. Most likely, the spiral manifests the Rossby waves with spiral crests which occur above the isolated underwater mountains (Rhines[29, 30]). In our case, the entire water layer with shallow water in the center and deep water at the wall of the tank, can be considered as that above a mountain. We will illustrate this effect using linear Rossby wave theory in Section C.

Fig. 3 shows the azimuthal velocity and relative vorticity in the Bc experiment. The warm water rises from the wire at the bottom to the surface forming a narrow upper layer initially aligned along the wire. This layer is lens-like in cross section and is unstable with respect to baroclinic/frontal instability. It rapidly breaks into small eddies of typical radius of approximately 1 cm. The size of the eddies is most likely determined by baroclinic radius of deformation. These eddies possess available potential energy which then slowly released into the system. During the forcing period strong vertical convective motions induced by hot wires is the main source of energy. The effect of this forcing can be described by the Stommel's[21] β-plume theory which was developed in application to hydrothermal vents in the ocean. When the forcing is switched off, the eddies' available potential energy becomes the main source which can sustain the flow for a very long time. Panels a and b of Fig. 3 show the flow during the forcing period. Comparison with the Bt experiment reveals that jets are more prominent in this case most likely because the eddies are much smaller and weaker than those induced by magnets. It is interesting to note that this observation agrees with the results of high-resolution numerical simulations by Scott and Drischel[31]. These authors reported that the onset of strong jets in the forced-dissipative barotropic β-plane turbulence occurs when the jet and eddy vorticity maxima are comparable, but jets weaken when the eddies are too strong. During the decay period the azimuthal velocity map (Fig. 3 c) shows a spiral pattern similar to that in Bt experiment. However, the relative vorticity map (Fig. 3 d) reveals that the eddies are much more "alive" in this experiment compared to the barotropic one. Note also that a cyclonic circulation develops in the tank which causes the asymmetry between the strength of the eastward and westward jets. This circulation is due to a gradual accumulation of warm water at the periphery of the tank. The circulation then occurs as suggested by the thermal wind balance with a radial gradient of density.

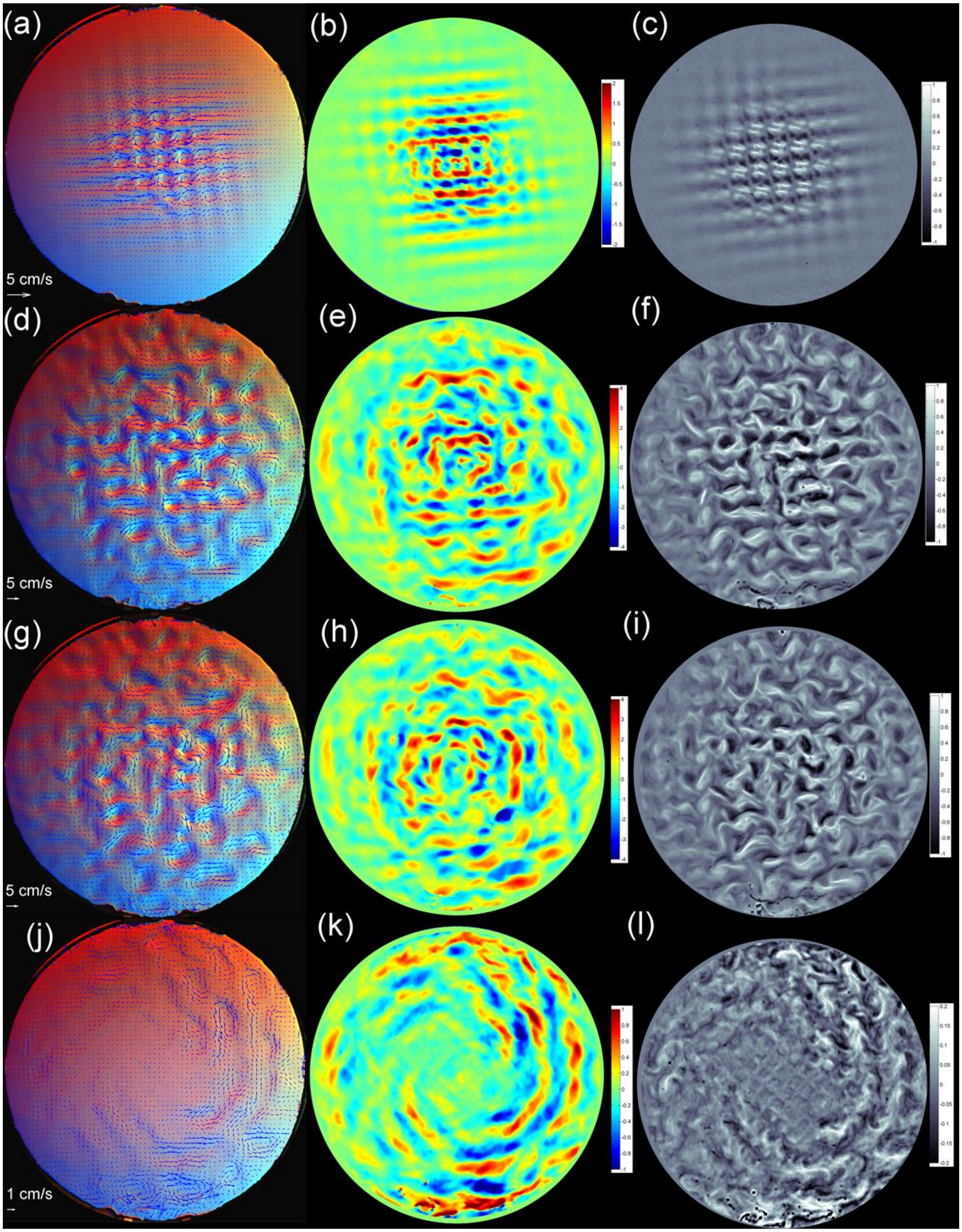

Fig. 2 (Color online) Evolution of the turbulent barotropic flow with β-effect visualized by AIV in the Bt experiment (enhanced online): initial period shortly after the forcing starts, $t$ = 2 s (a-c),

stationary forced turbulence, $t$ = 380 s (d-f) and decaying turbulence, $t$ = 2 s (g-i) and $t$ = 30 s after the forcing is stopped (j-l). The first column of panels shows velocity vectors superposed on color altimetry images. The velocity scale is given in the lower left corner of each panel. The second column shows the x-component of velocity, $u$, while the third column shows the dimensionless vorticity $\zeta/f_0$ varying from negative values (black, anticyclonic) to positive values (white, cyclonic).

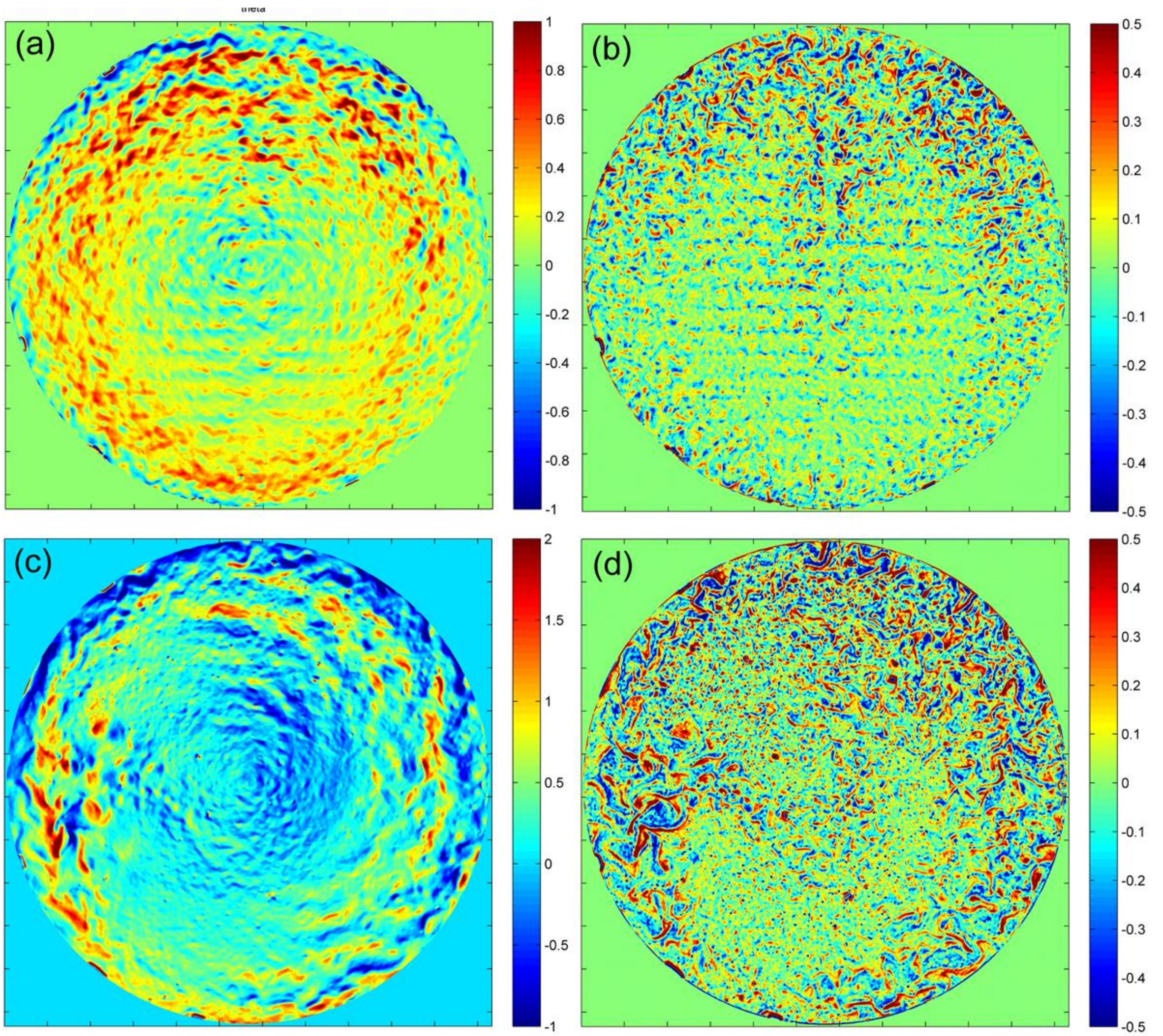

Fig. 3 (Color online) Baroclinic flow generated by thermal forcing in the Bc: forced turbulence, $t$ = 360 s (a, b) and decaying turbulence, $t$ = 40 s after the forcing is stopped (c, d). Panels a and d show the x-component of velocity, $u$; panels b and d show the dimensionless vorticity $\zeta/f_0$ .

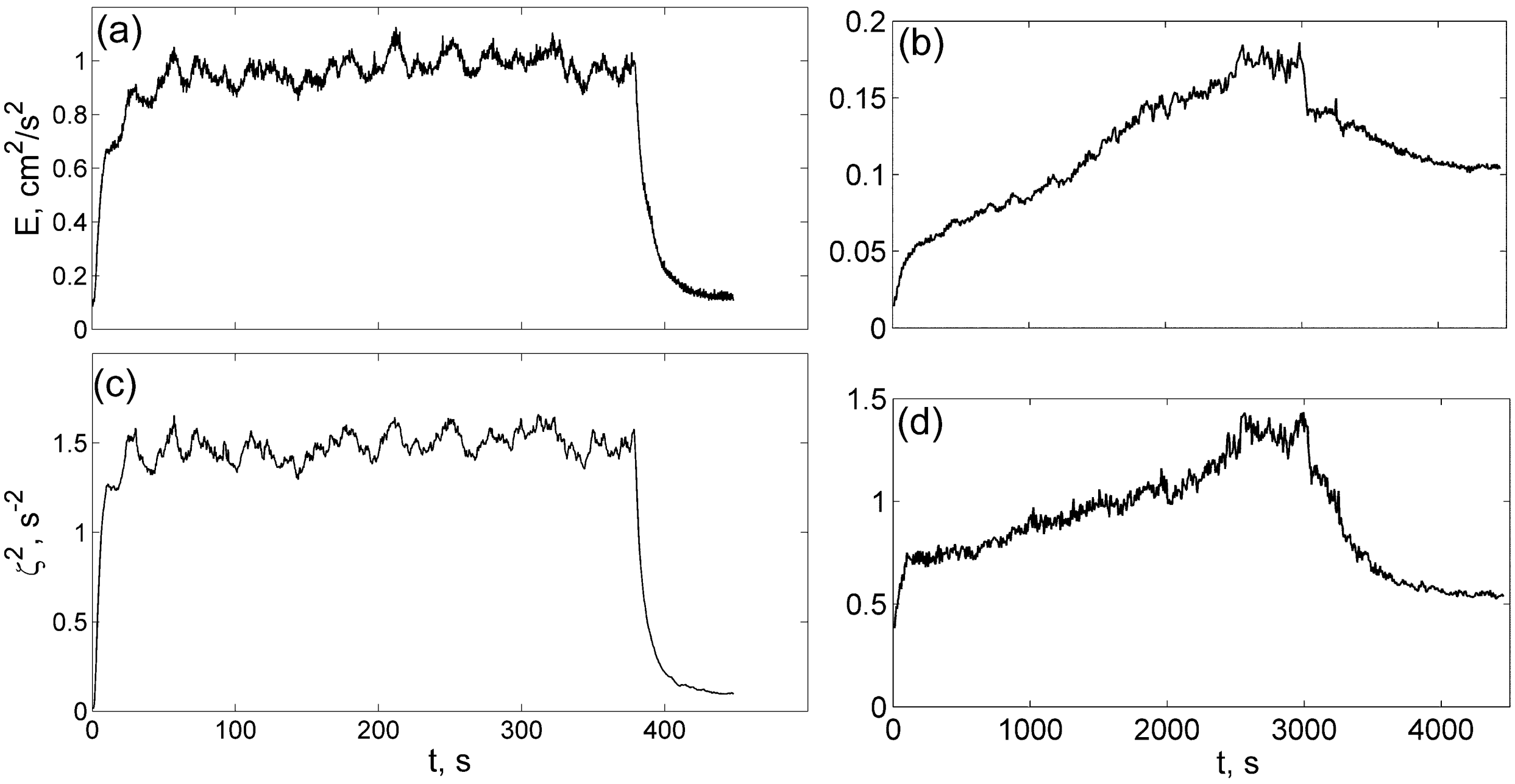

Fig. 4. Kinetic energy $E$ and enstrophy $\zeta^2$ versus time in the Bt (a, c) and Bc (b, d) experiments.

In order to establish that the flow becomes stationary during the forcing period we measured kinetic energy $E = \mathbf{U}^2/2$ and enstrophy $\zeta^2$ in both experiments. Time series of mean values of both quantities are shown in Fig. 4. In the Bt experiment both energy and enstrophy fluctuate around some stationary levels during the forcing period. In the Bc experiment both quantities continue to rise slowly over a long period of time until they reach approximately steady levels at $t$ = 2500 s. After the forcing stops both energy and enstrophy decay rapidly in the Bt experiment while in the Bc experiment the decay is much slower. The measurements of energy decay allow us to quantify dissipation in the experiments. Dissipation in the rotating fluid can be interpreted in terms of friction in the bottom Ekman layer and viscosity in the bulk of the layer. The rate of change of the mean energy in the system is given by

$$\frac{d\bar{E}}{dt} = -2\alpha\bar{E} - \nu\bar{\zeta}^2 \ , \qquad (4)$$

where bottom friction is parameterized by a linear term with linear drag coefficient $\alpha$ and $\nu$ is the kinematic viscosity of fluid. The coefficient $\alpha$ can then be determined as

$$\alpha = -\left[ d(\ln\bar{E})/dt + \nu\bar{\zeta}^2/\bar{E} \right]/2 .$$

Alternatively, a theoretical estimate $\alpha = (\Omega\nu)^{1/2}/H$ can be obtained from the bottom Ekman layer theory. The coefficient $\alpha$ is approximately equal to 0.05 $s^{-1}$ in the Bt experiment. The coefficient was measured immediately after the forcing stops using the energy and enstrophy data shown in Fig. 4 (a, c). Note that the contribution of "bulk" viscosity constitutes only 11% of the total rate of change of energy such that most of the energy is dissipated at the bottom. Ekman dissipation is not uniform across the domain; the system loses energy faster in the shallow area in the center than in the deeper layer at the periphery. Theory gives $\alpha$ = 0.06 $s^{-1}$ in the center and $\alpha$ = 0.01 $s^{-1}$ at the wall of the tank in the Bt experiment.

The presence of an energy source (baroclinic eddies) in the flow during the decay period in the Bc experiment makes it impossible to measure $\alpha$ using mean energy and enstrophy data. In this experiment we have to rely on the theoretical estimate only. It gives more uniform distribution with radius than that in the Bt experiment with $\alpha$ varying between 0.03 $s^{-1}$ in the center and 0.01 $s^{-1}$ at the wall. A bulk value (averaged over the area) is $\alpha$ = 0.02 $s^{-1}$.

An important control parameter in a forced-dissipative turbulence is the rate, $\varepsilon$, at which the forcing supplies energy to the system. In a stationary flow this energy is subsequently removed from the system by dissipation at the same rate. Assuming that immediately after the forcing stops the energy dissipation rate remains the same as it was during the forcing, $\varepsilon$, can be estimated as

$$\varepsilon = -\frac{d\bar{E}}{dt} \approx 2\alpha\bar{E} \ . \qquad (5)$$

The value of $\varepsilon$ in the Bt experiment is 0.1 cm$^2$/s$^3$ while in the Bc experiment $\varepsilon = 0.006$ cm$^2$/s$^3$.

### B. Energy spectra in wavenumber space

Further insight into the dynamics of the turbulent flow can be provided by an analysis of its spectral characteristics. For a Fourier decomposition it is convenient to use the Cartesian coordinates. We chose the local coordinate system which corresponds to the β-plane centered at the reference radius $r_0 = 2R/3$. The domain is of halfwidth $R/3$ such that we cut off the central part of the tank. The velocity vectors obtained by altimetry (in the global Cartesian coordinates) are mapped into the local coordinate system to obtain the velocity $\mathbf{u} = (u, v)$ where $u$ is the West-East and $v$ is the South-North component of the velocity vector.

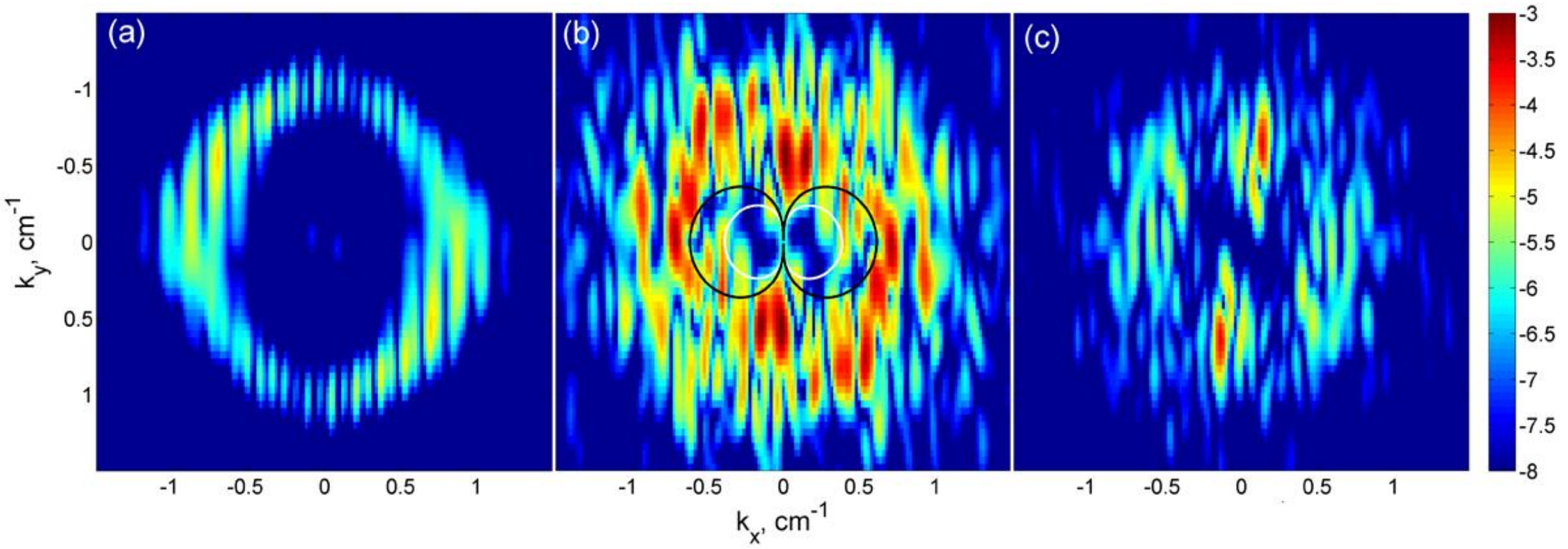

Fig. 5 (Color online) Evolution of the two-dimensional energy spectrum in the Bt experiment. The scale on the right shows the logarithm of energy in wavenumber space ($k_x$, $k_y$) at $t$ = 4 s (a), $t$ = 382 s (b) from the start of the experiment and at $t$ = 12 s (c) after the forcing is switched off. The white line in panel (b) shows $k_R$ given by Eq. (10) while the black line shows $k_\beta$ given by Eq. (11).

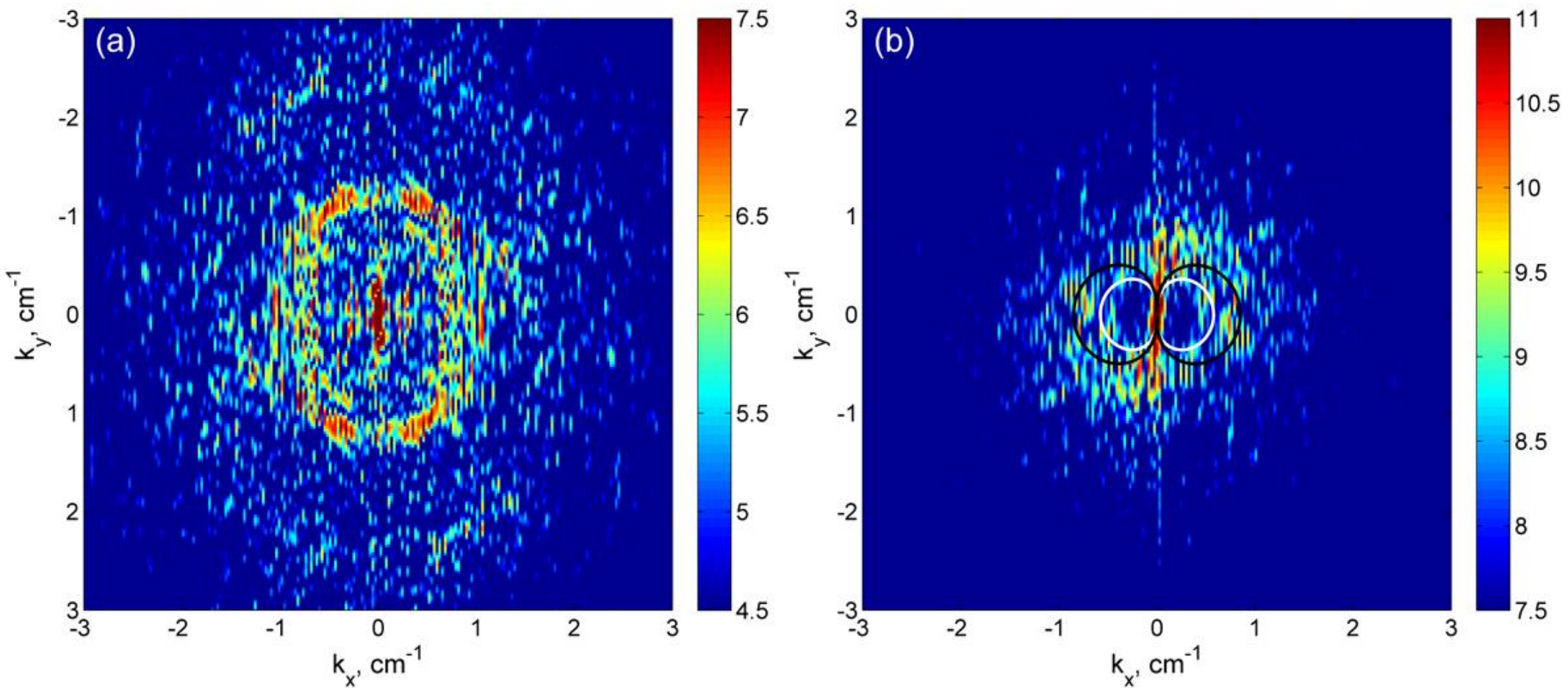

Fig. 6 (Color online) Two-dimensional energy spectrum in the Bc experiment in the beginning of the experiment at $t$ = 35 s (a) and at $t$ = 37 s (c) after the forcing is switched off. The white and black lines show $k_R$ and $k_\beta$ respectively.

Two-dimensional energy spectrum is given by

$$E(k_x, k_y) = \frac{1}{2} | \mathbf{u}(k_x, k_y) |^2 \ , \qquad (6)$$

where $(k_x, k_y)$ is the wavenumber and $\mathbf{u}(k_x, k_y)$ represents the discrete Fourier transform of the velocity vector. Figure 5 shows the evolution of the two-dimensional spectrum in the Bt experiment. The spectrum shortly after the forcing starts (Fig. 5 a) is in the form of a ring and is approximately isotropic. Note that since there are initially approximately 10 vortices of the same sign across the tank, the forcing wavelength is $\lambda_f = 2R/10 = 9$ cm (equal to twice the distance between the magnets). The forcing wavenumber is then $k_{xf} = k_{yf} = 2\pi/\lambda_f \approx 0.7$ cm$^{-1}$ which corresponds well with the size of the ring in Fig. 5 a. Panel b in Fig. 5 shows the spectrum of a stationary forced flow. An important feature of the spectrum is the concentration of energy at the line $k_x = 0$ which, in physical space, is an indication of zonal jets. Finally, panel c in Fig. 5 shows the spectra of the decaying flow when the system is no longer driven to isotropy by forcing. Strong anisotropy develops with most of the energy concentrated in zonal motions at $k_x = 0$ and $k_y \approx 0.7$ cm$^{-1}$. In the physical space, vortices decay while the zonal jets become more prominent.

Fig. 6 shows the two-dimensional spectrum in the Bc experiment. Initially (Fig. 6 a) the strongest signal is at approximately $k \approx 1$ cm$^{-1}$ which correspond to the spacing between the segments of the wire. The second, albeit weaker ring is visible at $k \approx 2.2$ cm$^{-1}$ . This wavenumber indicates the signature of small scale baroclinic eddies formed as a result of baroclinic instability. Eventually, energy cascades toward smaller wavenumbers as well as concentrates near the line $k_x = 0$. Fig. 6 b shows the spectrum after the forcing is switched off. The spectrum is qualitatively similar to that in the Bt experiment (Fig. 5 c).

Solid white and black lines in Figs. 5 b and 6 b show anisotropic Rhines wavenumber, $k_R$ and a wavenumber $k_\beta$ correspondently. These wavenumbers are determined using a "synchronization" condition for turbulence and Rossby waves which requires the frequency of the turbulent motions to be equal to the frequency of the waves. The frequency of the Rossby waves is determined by their dispersion relation

$$\omega(k_x, k_y) = \frac{\beta k_x}{k^2 + R_d^{-2}} \;, \qquad (7)$$

where $k^2 = k_x^2 + k_y^2$ and $R_d = (gH)^{1/2} / f_0$ is the radius of deformation. Turbulence, on the other hand, does not have a dispersion relation; the frequency of turbulent motions can be determined using different considerations. Rhines[12] used the root-mean-square (rms) velocity of vortices to obtain

$$\omega = U_{rms} k. \qquad (8)$$

Turbulence is assumed to be isotropic with wavenumber $k$ characterizing a vortex arrangement in the turbulent flow. Alternatively, the characteristic frequency of forced turbulence can be obtained using the energy rate $\varepsilon$ as a main control parameter. In the upscale energy cascade with spectral slope -5/3, the time required for a fluid parcel to move a distance $1/k$ is $\varepsilon^{-1/3}k^{-2/3}$ (see e.g. Vallis[32]) which gives

$$\omega = \varepsilon^{1/3}k^{2/3}. \qquad (9)$$

Equating turbulent frequencies (8) or (9) to the Rossby wave frequency (7) and solving for the wavenumber, we obtain the Rhines wavenumber

$$k_R = \left(\frac{\beta}{U_{rms}}\right)^{1/2} \cos^{1/2}\theta\,, \qquad (10)$$

and the $k_\beta$ wavenumber

$$k_\beta = \left(\frac{\beta^3}{\varepsilon}\right)^{1/5} \cos^{3/5}\theta\,. \qquad (11)$$

Here $\theta$ is the angle in the wavenumber plane such that $\cos\theta = k_x/k$. In the derivations of the wavenumbers we assumed $R_d$ to be large and ignored it in the dispersion relation (7).

Physically $k_\beta$ can be interpreted as a wavenumber where turbulence starts "feeling" the β-effect. A match between the "eddy frequency" and the Rossby wave frequency allows the turbulence to emit Rossby waves. The energy cascade, however, does not stop there and continues towards smaller wavenumbers, especially towards $k_x = 0$ (Vallis[32]). The Rhines wavenumber determines where the Rossby waves are emitted most effectively by turbulent motions of characteristic velocity $U_{rms}$. Note that this velocity is determined by the balance of the energy supply by the forcing and its withdrawal by dissipation. The Rhines wavenumber represents a boundary which divides small-scale turbulence and large-scale wave dominated motions. The lines in Figs. 5 b and 6 b are calculated using parameters measured in the experiments: $U_{rms} = (2\bar{E})^{1/2}$ and $\varepsilon = 2\alpha\bar{E}$ . It is clear that the space inside the dumbbells, where the Rossby waves dominate, contains reduced energy.

The lines showing anisotropic wavenumbers $k_R$ and $k_\beta$ in Figs. 5 b and 6 b represent certain frequencies selected using the criteria specified by equations (8) and (9). However, turbulent motions emit waves with the entire spectrum of frequencies. Rossby wave emission by a localized perturbation is described by the β-plume theory (see e.g. Refs (21, 22)). Afanasyev[36] employed the β-plume approach to show the evolution of two-dimensional spectrum of turbulence towards the anisotropic state similar to that observed in in Figs. 5 b and 6 b. The emission of Rossby waves occurs due to a forcing in the RHS of a linearized quasi-geostrophic equation

$$\frac{\partial}{\partial t}\left(\nabla^2 - R_d^{-2}\right)\eta + \beta\frac{\partial \eta}{\partial x} = F \ , \tag{12}$$

where $\eta$ is the surface elevation. The forcing is isotropic with a known spectrum. Note that an isotropic small-scale turbulence with the developed (due to nonlinear triad interactions) -5/3 type energy cascade can be considered as forcing for this purpose. A solution for the Fourier transform of the surface elevation $\eta$ can be easily obtained as

$$\eta(k_x, k_y) = \frac{iF(k_x, k_y)}{k_x\beta}\left[\exp(-i\omega t) - 1\right], \tag{13}$$

where $F(k_x, k_y)$ is the forcing and $\omega$ is the frequency of the Rossby waves given by the dispersion relation (7). Geostrophic velocity is obtained from the surface elevation as

$$\mathbf{u}_g(k_x, k_y) = \frac{g}{f_0}(ik_y, -ik_x)\eta(k_x, k_y) \ , \tag{14}$$

and the two-dimensional energy spectrum is then calculated using Eq. (6). Fig. 7 shows the evolution of the spectra generated by an isotropic Gaussian forcing.

$$F(k_x, k_y) = \exp\left( -\frac{(k - k_f)^2}{0.04} \right), \tag{15}$$

where $k_f$ = 1 cm$^{-1}$ is the forcing wavenumber. The ring-like Gaussian forcing approximately models the forcing applied by the magnets in our experiment. The dimensional control parameters used to calculate the spectrum shown in Fig. 7 are chosen to have the same values as in the Bt experiment. Fig. 7 shows that the energy spectrum develops from the initially isotropic spectrum towards a strongly anisotropic one where the energy is concentrated at the $k_x$ = 0 line similar to that in the experiment (in Figs. 5 b and 6 b).

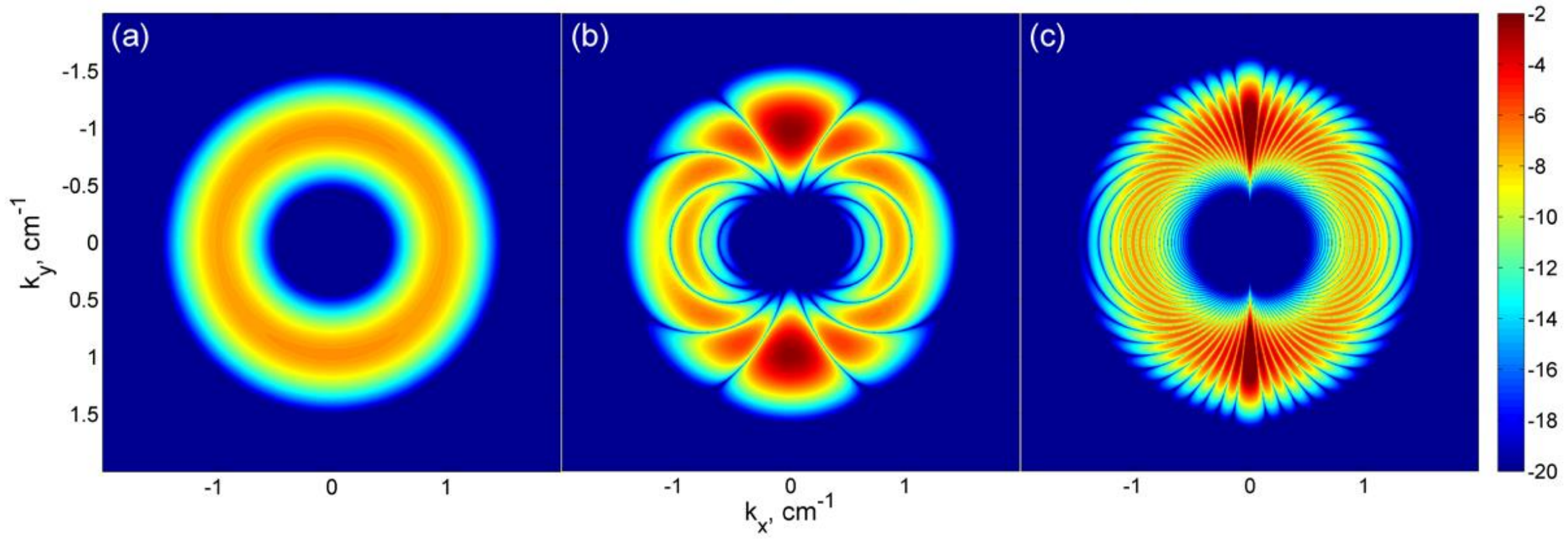

Fig. 7. The evolution of the energy spectrum given by the theoretical solution (13). The scale on the right shows the logarithm of energy in wavenumber space ($k_x$, $k_y$) at $t$ = 10 s (a), $t$ = 100 s (b) and at $t$ = 500 s (c). The dimensional parameters are $k_f$ = 1 cm$^{-1}$, $f_0$ = 4.64 s$^{-1}$, $\beta$ = 0.2 cm$^{-1}$s$^{-1}$ and $R_d$ = 18 cm.

One-dimensional energy spectrum is an important characteristic of a turbulent flow which can be easily obtained from the two-dimensional spectrum such as that shown in Figs. 5 and 6. The one-dimensional (isotropic) spectrum is defined as

$$E(k) = 2\pi k \bar{E}(k_x, k_y), \qquad (15)$$

where the two-dimensional spectrum $E(k_x, k_y)$ is averaged over angle $\theta$ in wavenumber domain. In anisotropic β-plane turbulence, a zonal spectrum can be defined as:

$$E_Z = E(k_x = 0, k_y). \qquad (16)$$

Theory predicts that if the certain criteria are satisfied the flow has a steep zonal spectrum (Rhines[12], Sukoriansky et al.[13] )

$$E_Z(k_y) \sim \beta^2 k_y^{-5}, \qquad (17)$$

while the non-zonal spectrum is expected to have a slope of -5/3 as predicted by Kolmogorov-Kraichnan theory for two-dimensional turbulence.

Fig. 8 shows both the isotropic spectrum $E(k)$ and zonal spectrum $E_Z(k_y)$ in the Bt experiment while Fig. 9 shows the spectra in the Bc experiment. In both experiments, in forced-dissipative regime the isotropic spectrum shows a good agreement with the predictions of the two-dimensional turbulence theory. The enstrophy and energy ranges are separated by a forcing wavenumber. In the Bt experiment the isotropic forcing wavenumber is $k_f = \sqrt{2} k_{xf} \approx 1$ cm$^{-1}$ (arrow in Fig. 8). In the Bc experiment the forcing wavenumber corresponds to the size of the baroclinic eddies and is more difficult to pinpoint precisely but it is clearly smaller than the forcing wavenumber in the Bt experiment. In the enstrophy range ($k > k_f$) the slope of the energy spectra is close to -3 in both experiments. The slope in the energy range ($k < k_f$) is close to -5/3 in both experiments. Note, however, that the energy range in the Bt experiment is quite limited. The RSW experiments by Afanasyev and Craig[6] showed that the finite radius of deformation can be a limiting factor for the cascade (see also Danilov and Gurarie[1]). In the ring-shaped domain we use for measuring spectra the values of $R_d$ vary between 13 cm and 26 cm (bulk value 20 cm) and between 15 cm and 23.5 cm (bulk value 21 cm) in the Bt and Bc experiments respectively. These values are similar to those in Ref. (6) and so are the lowest limits of the energy range.

The zonal spectra $E_Z(k_y)$ peak at low wavenumbers and decline, albeit somewhat flatter than -5, at larger wavenumbers. Compensated energy spectra in the form $E(k)\varepsilon^{-2/3}k^{5/3}$ are shown in panel c of Figs. 8 and 9. The compensated spectra in both experiments exhibit plateaus at the energy range that allows us to determine the value of the Kolmogorov constant, $C \approx 6$ and 5 in the Bt and Bc experiments respectively. Compensated zonal spectra, $E_Z(k_y)\beta^{-2} k_y^5$ (Fig. 8 d and Fig. 9 d), however, vary with $k_y$ and do not allow an accurate measurement of a constant in the theoretically predicted -5 law.

The spectra in Fig. 8 b and Fig. 9 b were measured after the forcing was switched off and the flows were decaying. In the Bt experiment the levels of energy decrease; the shape of the zonal spectrum remains almost the same while the isotropic total energy spectrum loses its -5/3 range and exhibits a uniform slope of approximately -3 from the low wavenumber cut-off at $k \approx 0.5$ cm$^{-1}$ towards larger wavenumbers. In contrast, in the Bc experiment the flow is still driven by the baroclinic energy source at relatively large wavenumbers even when the direct forcing is switched off. The -5/3 range remains and is even wider than that in Fig. 9 a. Fig. 10 shows the evolution of the energy spectra $E(k)$ after the forcing is switched off in each experiment. The levels of energy drop and the peaks of spectra shift to lower wavenumbers. To characterize the shift of the spectral peak it is useful to consider the energy-weighted mean wavenumber $k_E$ defined as

$$k_E = \int_0^\infty kE(k)dk \,/ \int_0^\infty E(k)dk \,.$$

Wavenumber $k_E$ provides a measure of the energy containing scale. Evolution of $k_E$ just before and after the forcing was stopped is shown in Fig. 10 c, d. In both Bt and Bc experiments the energy-weighted mean wavenumbers stay at some constant level during the forcing and then drop to a lower level when the forcing is stopped. Similar behaviour was also observed in previous EM experiments by Afanasyev and Wells[23].

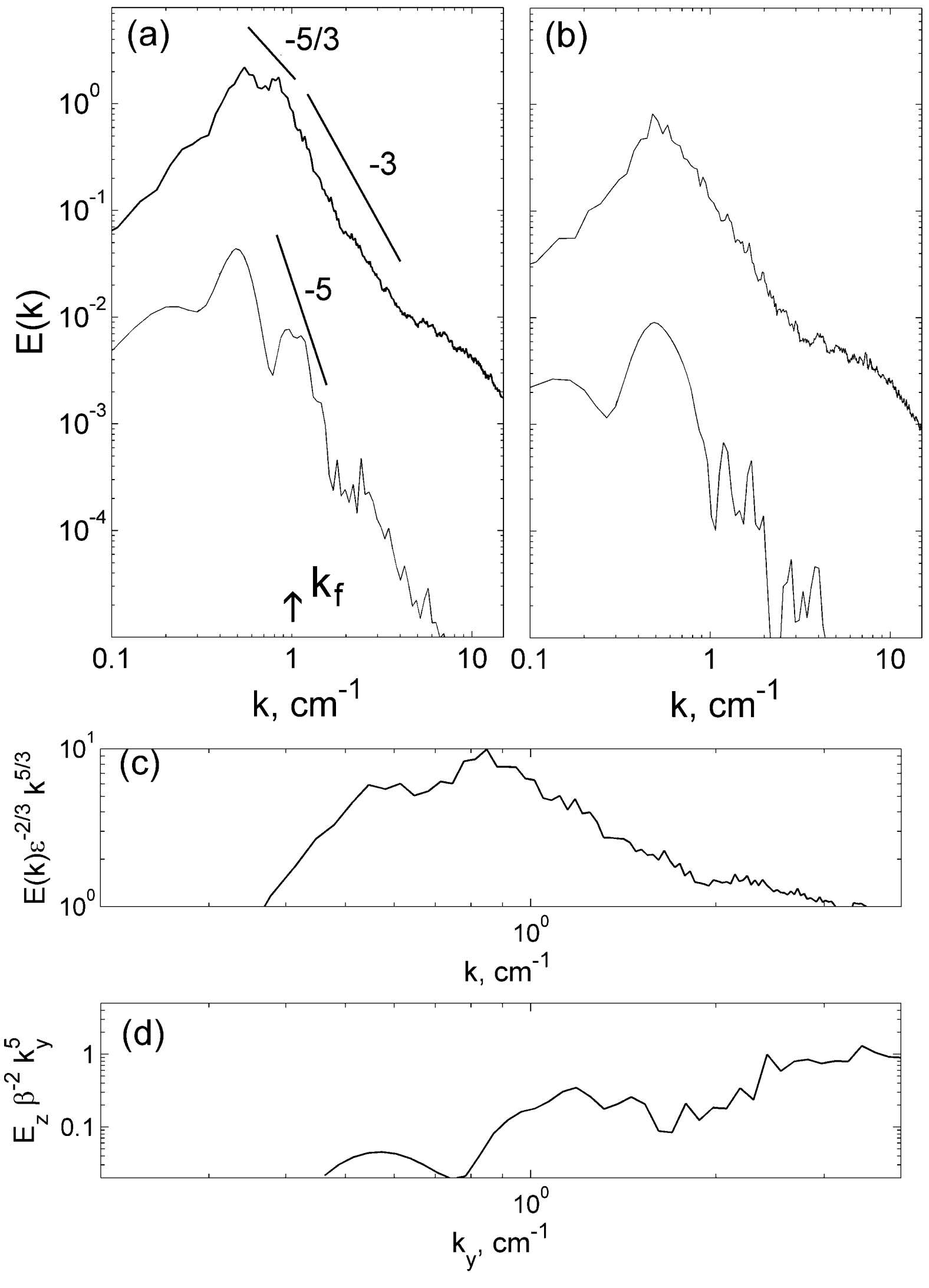

Fig. 8. The one-dimensional energy spectra in forced (a) and in decaying turbulence (b): energy $E(k)$ (upper curve) and zonal energy $E_Z(k_y)$ (lower curve). Arrows indicate the forcing wavenumber $k_f$.

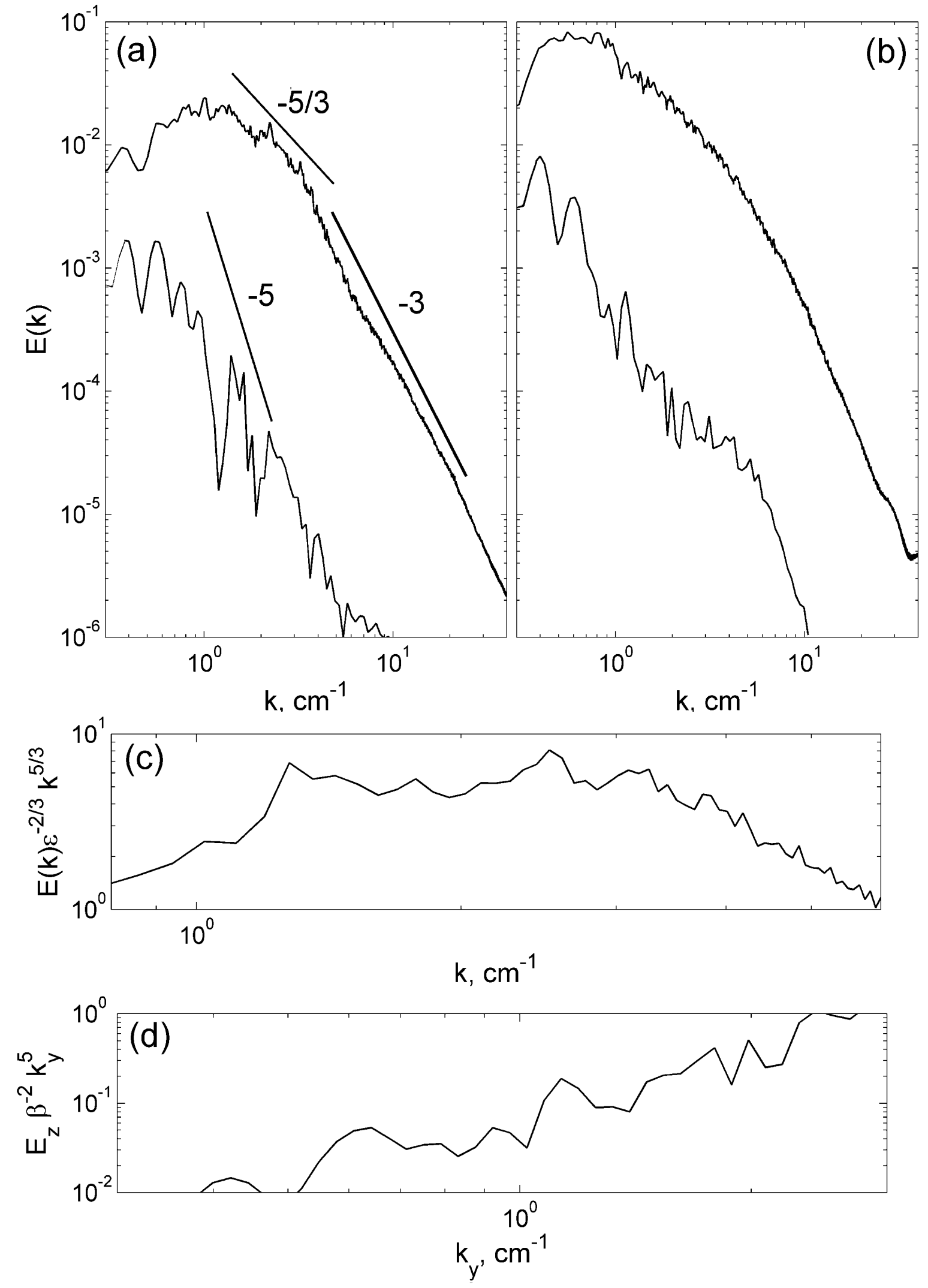

Fig. 9. The one-dimensional energy spectra in forced (a) and in decaying turbulence (b): energy $E(k)$ (upper curve) and zonal energy $E_Z(k_y)$ (lower curve). Arrows indicate the forcing wavenumber $k_f$.

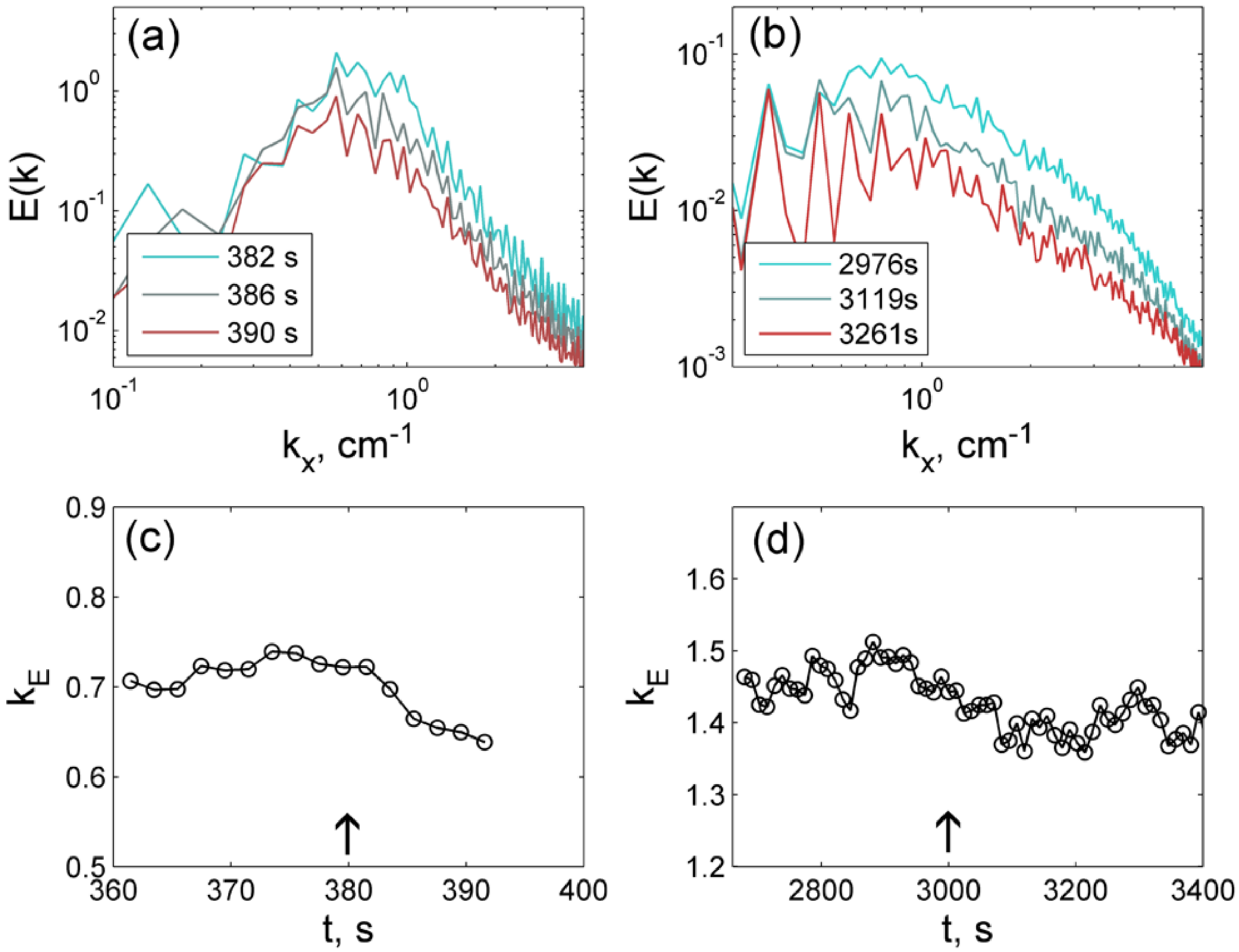

Fig. 10. Evolution of the one-dimensional energy spectra (a, b) and the energy-weighted mean wavenumber $k_E$ (c, d) in decaying turbulence in the Bt (left column) and Bt (right column) experiments. Arrows indicate time when the forcing is switched off in each experiment.

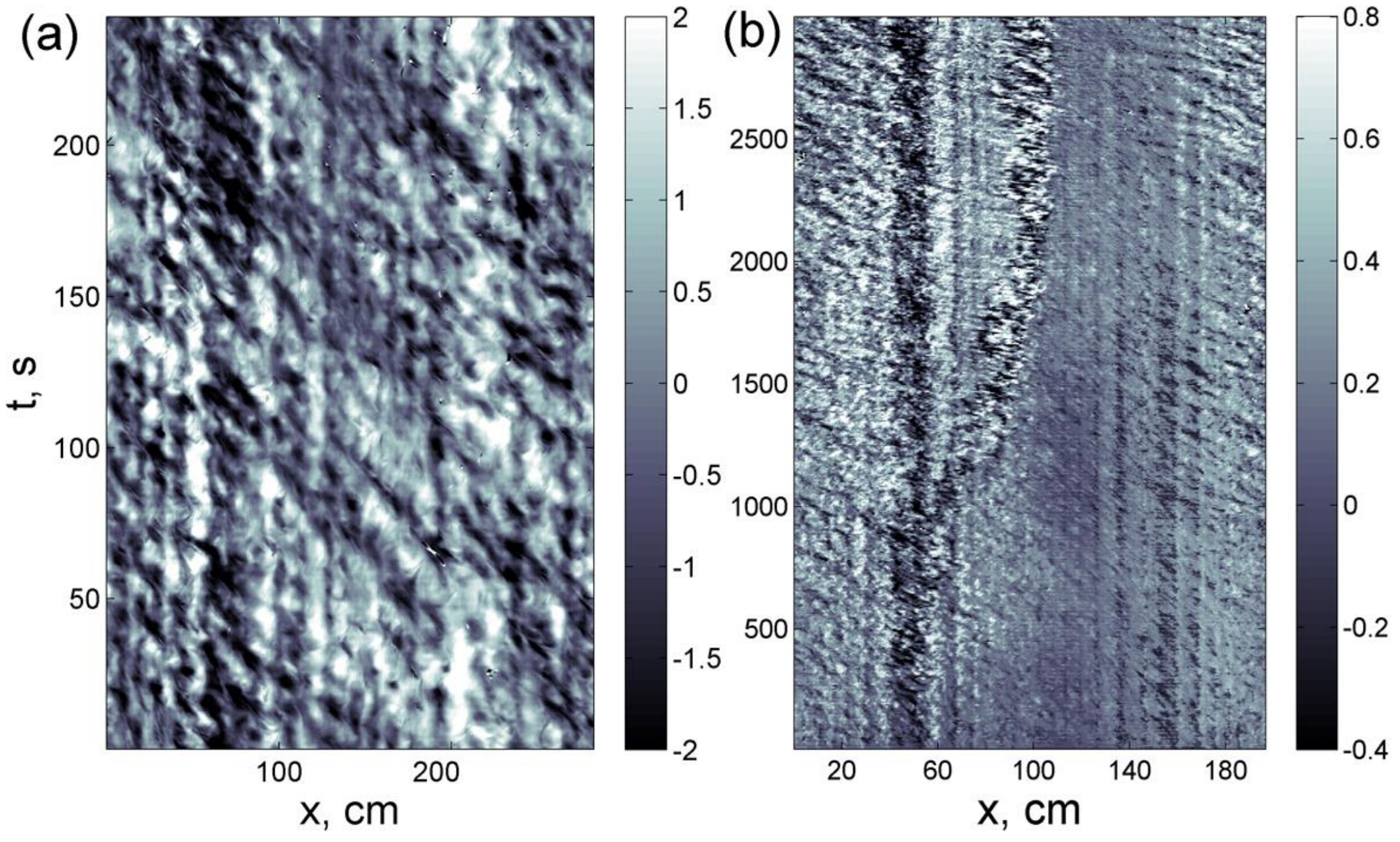

Fig. 11. Velocity $u$ measured along a circle of radius $r_0$ equal to two thirds of the radius of the domain at different times in the Bt (a) and Bc (b) experiments. The grayscale shows $u$ in cm/s.

### C. Frequency-wavenumber space and Rossby waves

In his pioneering paper on β-plane turbulence Rhines[12] predicted that Rossby waves should dominate at low wavenumbers below the wavenumber $k_R$. In what follows we shall provide experimental evidence of the Rossby waves in our experiments. It is not easy to identify waves in a turbulent flow. One way to do that is to use the fact that Rossby waves have a distinct dispersion relation (7). The dispersion relation shows in particular a well-known property, namely the westward phase propagation of the Rossby waves. Fig. 11 a, b shows the x-component of velocity, $u$, measured along a circle of radius $r_0 = 2R/3$ and $r_0 = 2R_t/3$ at different times during the Bt and Bc experiments respectively. The Hovmöller plot in Fig. 11 shows $u$ in greyscale with distance along the circle in the horizontal and time in the vertical axes. Human eye is good at identifying patterns and it is easy to see a pattern of features aligned in oblique straight lines in these plots. These features are perturbations of the velocity $u$ and they propagate in the negative x-direction, to the west. The slope of the lines then gives the phase velocity of the waves. Similar measurements performed at different radii $r_0$ show that the Rossby waves present everywhere in the tank.

The dispersion relation (7) can be used in a more direct way to identify the spectral signature of the Rossby waves. Fig. 12 shows the energy spectra in a frequency-wavenumber domain in both Bt and Bc experiments. To calculate the spectra we measured the velocity ($u$, $v$) along the circle $r_0$ during a long time period when the flow is forced and then performed the Fourier transform in time and in distance, $x$, along the circle. The resultant spectra show some interesting features of the flow. In the Bt experiment where the forcing was relatively strong and fixed in space, high energy motions are mostly concentrated at almost zero frequencies with "hot" spots at $k_x \approx \pm 0.8$ cm$^{-1}$, close to forcing wavenumber, and at $k_x = 0$. These high-energy, almost stationary motions radiate towards higher frequencies. In contrast, zero-frequency motions are not so obvious in the Bc experiment and main energy is concentrated in a plume extending to higher frequencies. The distribution of spectral energy at higher frequencies is clearly asymmetric with respect to the wavenumber $k_x$ in both experiments. Most of the energy is at negative $k_x$ which indicates that westward propagating waves dominate in the flow. The white lines in both panels in Fig. 12 show the Rossby waves dispersion relation (7) calculated with different values of $k_y$. The upper curves are calculated with $k_y = 2\pi/2R = 0.07$ cm$^{-1}$(Bt experiment) and $k_y = 0.06$ cm$^{-1}$ (Bc experiment) which correspond to largest possible wavelengths that fit across the tank. These curves provide an upper limit for the frequencies of the Rossby waves. The lower curves are calculated with the wavenumber $k_y = 0.7$ cm$^{-1}$ equal to the forcing wavenumber. Thus, most of the Rossby waves that can be excited in this system should be in between these two lines. According to Rhines[12] waves are excited most effectively at frequencies given by the turbulence "dispersion relation" Eq. (8). The straight white lines in Fig. 12 show Eq. (8); there is indeed some concentration of energy along these lines.

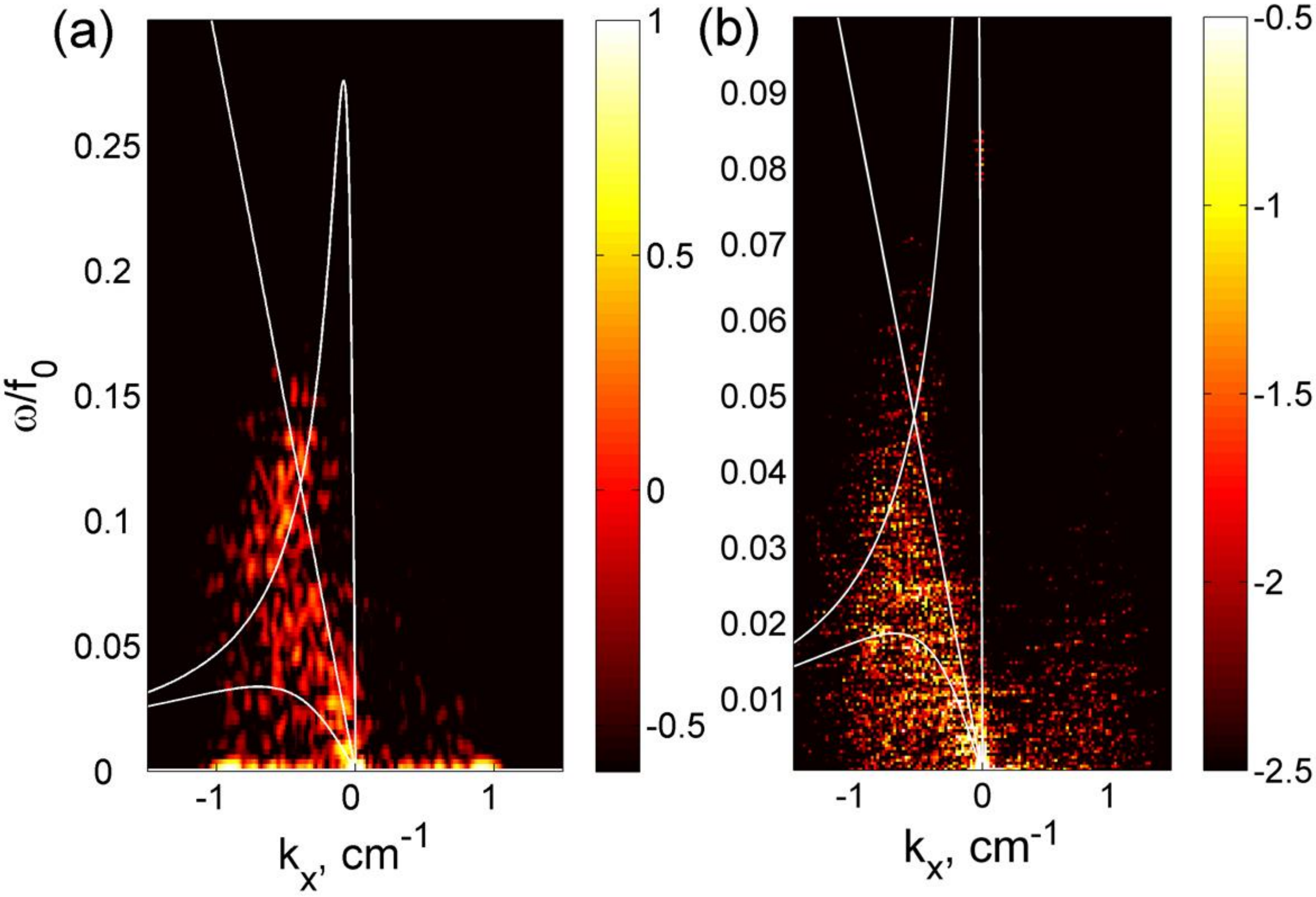

Fig. 12 (Color online) The energy spectrum in the frequency-wavenumber space. The scale shows the logarithm of energy. The frequency is normalized by the Coriolis parameter $f_0$. White curves show the Rossby wave dispersion relation (7) with $k_y$ = 0.07 and $R_d$ = 20 cm (a), and $k_y$ = 0.06 and $R_d$ = 19 cm (b) (upper curves); and with $k_y$ = 0.7 and $R_d$ = 20 cm (a), $k_y$ = 0.7 and $R_d$ = 19 cm (b) (lower curves). The solid straight lines shows turbulent frequency given by Eq. (8) with $U_{rms}$ = 1.34 cm/s (a) and $U_{rms}$ = 0.4 cm/s (b).

Let us return now to the spiral pattern of jets that we pointed out earlier in Fig. 2 k and Fig. 3 c. The spiral patterns were observed in the decaying flows. The energy is dissipated faster in the center of the domain where water is shallow. Therefore, we can expect propagation of energy from the periphery where the turbulence is still active towards the center. Turbulent eddies excite Rossby waves which can be described by linear theory. Here, however, we need to use a polar domain rather than a regular β-plane. Linearised quasi-geostrophic equation similar to Eq. (12) but in polar coordinates ($r$, $\theta$) can be written as

$$\frac{\partial}{\partial t}\left(\nabla^2 - R_d^{-2}\right)\eta + 2\gamma\frac{\partial\eta}{\partial\theta} = 0\,. \tag{18}$$

Here we use a so-called polar β-plane such that either the Coriolis parameter varies quadratically with distance from the pole on a rotating planet, $f = f_0 - \gamma r^2$, or water depth varies quadratically with radius in the laboratory setting. In the latter case the coefficient $\gamma$ is defined as

$$\gamma = \frac{2\Omega\left(\Omega^2/2g + c_b\right)}{H_0 - H_b}\,. \tag{19}$$

Looking for the solution of Eq. (18) in the form $\eta = \Phi(\kappa r)\exp(im\theta - i\omega t)$ one obtains a dispersion relation for the Rossby wave on a polar β-plane (Rhines[15])

$$\omega = \frac{2\gamma m}{\kappa^2 + R_d^{-2}}\;. \tag{20}$$

Radial dependence is given by either the Bessel function $\Phi(\kappa r) = J_m(\kappa r)$ or the Hankel functions $H_m^{(1)}(\kappa r)$ and $H_m^{(2)}(\kappa r)$. The former case corresponds to standing waves while the latter case describes waves propagating either inwards or outwards. Fig. 13 shows the Hankel wave with spiral wavecrests which corresponds to wavemode $m$ = 8, $\kappa$ = 0.73 cm$^{-1}$. The particular values of the wavenumbers were chosen to match the pattern with that observed in the Bc experiment (Fig. 3 c). Thus, this simple linear analysis allows one to see where the experiments are in the relevant wavenumber space.

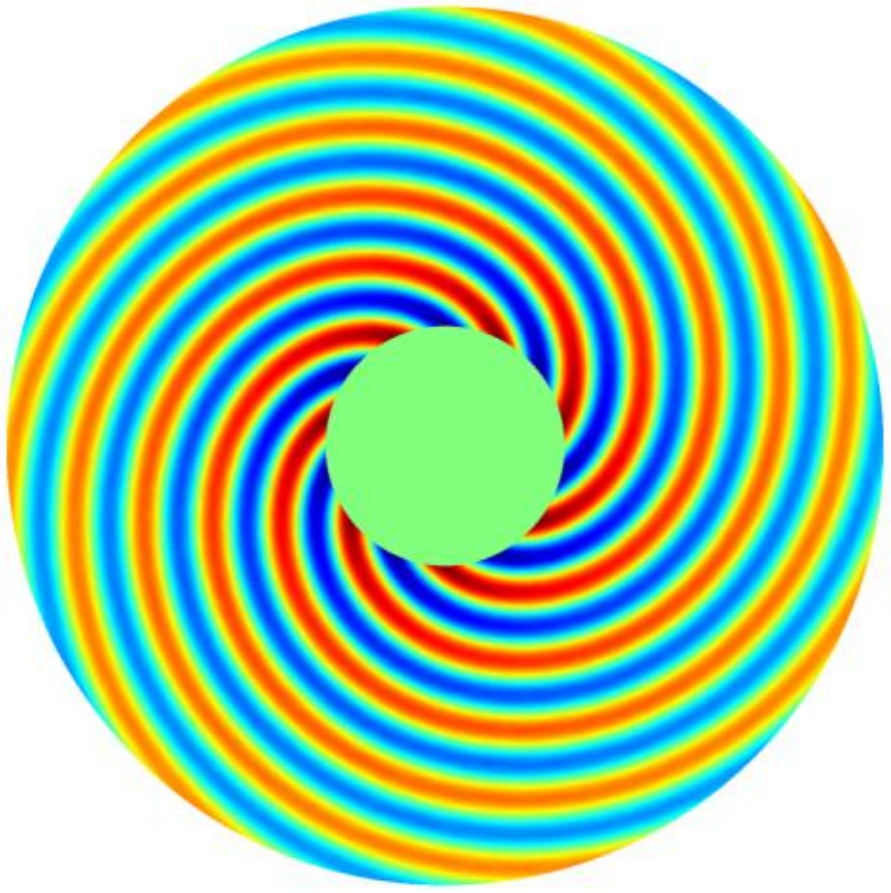

Fig. 13. Rossby wave with spiral wavecrests propagating towards the center of the domain and represented by a solution in the form of the Hankel function.

**DISCUSSION AND CONCLUSIONS**

In this work, we have shown experimental evidence that zonal jets occur both in forced-dissipative and in decaying turbulent flows on the topographic β-plane. The jets are latent in the forced-dissipative case if the forcing is strong and eddies dominate; the jets in our barotropic (Bt) experiment are similar in this respect to oceanic jets (also called “striations”) which can only be revealed by filtering and time-averaging of the velocity fields[9, 10]. The jets are stronger, relative to the eddies, in the baroclinic (Bc) experiment where the forcing was weaker. The jets become prominent in the decaying regime in both experiments, when the flow is no longer driven to isotropy by strong forcing.

Numerical simulations of β-plane turbulence by Galperin et al.[33], Galperin et al.[34], Sukoriansky et al.[13] and Scott and Dritschel[31] showed that the ratio $R_\beta = k_\beta/k_R$ controls the regime of the flow. When $R_\beta$ is large enough ($R_\beta > 2$), the flow is in a so-called zonostrophic regime when the jets are a dominant feature of the flow. The jets on gas giants satisfy the criterion of zonostrophy. When $R_\beta$ is less than approximately 1.5, the flow is in a viscous regime. A regime between the viscous and zonostrophic regimes ($1.5 < R_\beta < 2$) is transitional. The jets in the Earth's oceans seem to be in the transitional regime and so are the jets in our Bt experiments where $R_\beta \approx 1.7$.

The measurements of the energy spectrum in the wavenumber space demonstrated that the spectrum evolves from initially almost isotropic one to an anisotropic spectrum of a typical dumbbell form which was previously observed in numerical simulations (Vallis and Maltrud[35]). The evolution of the experimental spectrum is in a qualitative agreement with that predicted by a β-plume theory[36]. The energy cascades towards $k_x = 0$ and nonzero $k_y$ close to the forcing wavenumber.

Investigations of the energy spectra in the frequency-wavenumber domain revealed an asymmetry of the spectra with respect to the sign of the zonal wavenumber $k_x$. This asymmetry can be explained by the westward propagation of the Rossby waves. Some concentration of waves energy was observed along the line $\omega = U_{rms}k$ as predicted by Rhines[12].

The asymmetric frequency-wavenumber energy spectra, which are qualitatively similar to those observed in our experiments, were also observed in the middle latitude ocean by Wunsch[37]. In the ocean, the spectra also indicated a predominantly westward propagation of perturbations. A significant energy was distributed along the so-called `non-dispersive' line given by $\omega = \beta R_d^2 k$. This line was tangent to the first-mode baroclinic Rossby wave dispersion curve. Note that in our experiments the Rossby waves were barotropic, so the "non-dispersion" line represented a barotropic Rossby wave speed for the longest spatial scale (the maximum wave speed allowed by local water depth). However, in our case, shorter waves with wavenumbers close to the forcing wavenumber were more prominent than long non-dispersive waves.

In this work we do not distinguish between jets and almost zero-frequency Rossby waves with nearly zonal wavecrests. In Fig. 2 j-l and Fig. 3 c the (slightly spiralling) jets can be directly associated with Rossby waves emitted by eddies in the decaying flows. Moreover, we suggest that β-plume mechanism or emission of low-frequency waves by perturbations is the underlying mechanism in jets formation by β-plane turbulence. This notion is illustrated by Fig. 7 showing the evolution of the two-dimensional spectrum towards zonal motions solely by linear wave emission. Further sustaining the jets by turbulence will perhaps involve nonlinear mechanisms where a self-organization of the jet-eddies system is important and results in fluxes of momentum from eddies to jets.

## ACKNOWLEDGMENTS

The authors are grateful to Alexander Slavin for his help with one of the experiments. YDA is supported by the Natural Sciences and Engineering Research Council of Canada.